\documentclass[aps,prb,twocolumn,floatfix,
showpacs,
amssymb]{revtex4}
\usepackage{graphicx}

\usepackage{dcolumn}
\usepackage{amsmath}

\begin{document}

\preprint{decoherence9.tex... 7/2/02}

\title{Quantum dynamics, dissipation, and asymmetry effects in
quantum dot arrays}

\author{F. Rojas}
\author{E. Cota}
\affiliation{Centro de Ciencias de la Materia Condensada -- UNAM,
Ensenada, Baja California, M\'exico  22800}

\author{S. E. Ulloa} \affiliation{Department of Physics and
Astronomy, and Nanoscale and Quantum Phenomena Institute, Ohio
University, Athens, OH 45701-2979}

\begin{abstract}

We study the role of dissipation and structural defects on the
time evolution of quantum dot arrays with mobile charges under
external driving fields.  These structures, proposed as quantum
dot cellular automata, exhibit interesting quantum dynamics which
we describe in terms of equations of motion for the density
matrix. Using an open system approach, we study the role of
asymmetries and the microscopic electron-phonon interaction on the
general dynamical behavior of the charge distribution
(polarization) of such systems.  We find that the system response
to the driving field is improved at low temperatures (and/or weak
phonon coupling), before deteriorating as temperature and
asymmetry increase. In addition to the study of the time evolution
of polarization, we explore the linear entropy of the system in
order to gain further insights into the competition between
coherent evolution and dissipative processes.
\end{abstract}

\date{7/2/02}
\maketitle

\section{Introduction}

The size reduction of devices in microelectronics, and the
fundamental limitations encountered or anticipated due to quantum
mechanical effects, have turned attention to understanding
nanometer scale structures.  Systems where intrinsic quantum
mechanical effects can play an important role and could even be
used to advantage are under intense scrutiny. One recent prominent
example of such systems are quantum dot cellular automata.

\begin{figure}
\includegraphics[width=3.6in]{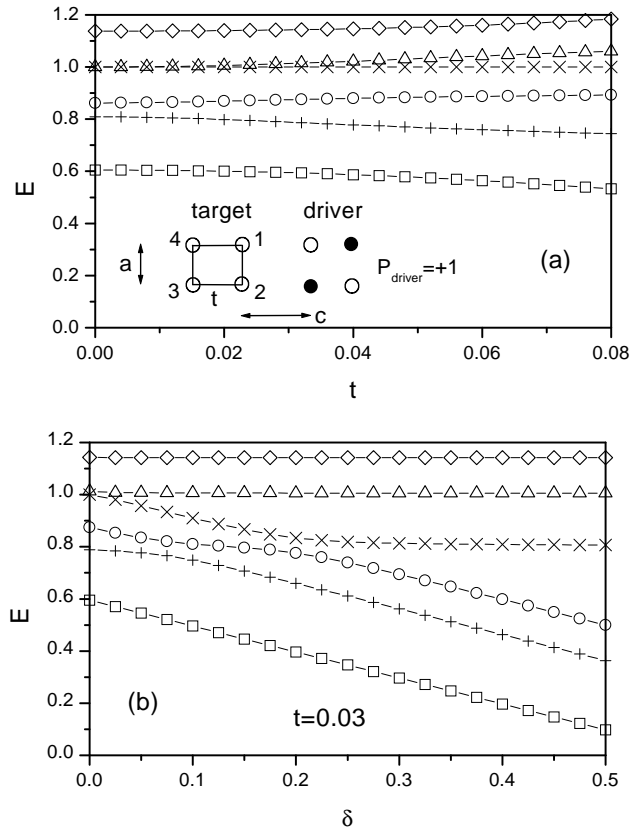}
\caption{(a) Energy level structure vs.\ $t$ for a target cell
with driver at +1. Energy in units of $V \simeq 1$ meV, and
$\delta=0$. Sketch at bottom: model QCA with four quantum dots per
cell. Two charges are allowed to tunnel between nearest neighbors,
with probability $t$. Driver cell on right indicates a state with
polarization $P_{driver}=+1$. (b) Level structure vs.\ dot size
asymmetry $\delta$.  Notice anticrossings for $\delta \simeq 0.1$
and 0.2 in excited states. Separation to driver, $c=a$ in both
panels.} \label{modelo}
\end{figure}

The original concept of ``cellular automata" was introduced as
$n$-dimensional arrays containing finite amplitude (information)
per cell (or site) and connected to one another according to
certain rules.  This simple definition, however, governs the
evolution of the entire array in rather complex ways.
\cite{wolfram} {\em Quantum} cellular automata are a natural
generalization of this concept, where each site contains a quantum
mechanical probability amplitude.\cite{gross} In 1993, Lent and
coworkers, \cite{lent93} proposed the use of cellular automata
architectures, composed of nanometer-scaled quantum devices
(quantum dots) coupled through carefully chosen hopping and
Coulomb interactions, to encode {\em classical} binary information
in the different charge configurations of the system. The typical
basic element in these {\em quantum-dot cellular automata} (QCA)
is a cell consisting of four quantum dots (QD) located at the
vertices of a square and connected via tunneling barriers to their
neighbors (see Fig.\ \ref{modelo}a). For two mobile electrons in
each four-dot cell, Coulomb repulsion between the electrons causes
the charge in the cell, in the ground state, to align along either
of the two diagonals. The symmetrical arrangement of the system
means that these two ``polarization'' states are degenerate and
can be used to represent logic $0$ and $1$ as the bits in this
system. The degeneracy is split by an appropriately designed
driver field which allows control of the polarization of the
ground state (and typically implemented as a second identical cell
with externally controlled polarization). Tunneling barriers
between dots are designed in such a way that intra-cell tunneling
is possible, but {\em inter}-cell tunneling is not, and only
Coulomb interaction from cell to cell is possible. Elemental cell
designs have been experimentally realized using metal islands, and
logic operations AND, OR and NOT have been implemented using
arrays of such cells suitably arranged. \cite{orlov, amlaniLOG,
amlani} Also, a clever implementation of QCA using magnetic
elements has been recently reported. \cite{Cowburn00}

Although originally proposed as `classical' and near
dissipationless bit operators, QCA could in principle also be used
to implement a quantum computer, as discussed recently,
\cite{Toth-Lent} if the entire array system is coherent enough
(and for sufficiently long times). In either the classical or
quantum operation mode, QCA computers (as well as any other)
should keep errors to a manageable limit. Of special significance
in these systems, decoherence and dissipation are external sources
of `error' originating in the coupling between the system and its
surroundings. Other aspects that have to be considered as possible
sources of loss of control or desirable operation are
imperfections in the dot fabrication, stray charged impurities,
and the role of switching fields or drivers. It is then of
interest to study the dynamical evolution of the QCA and monitor
the charge density distribution over the basic cell elements under
the influence of a driver cell, with emphasis on the effects of
imperfections and dissipation. The charge configuration is
conveniently monitored through Lent's cell {\em polarization},
\cite{lent93} which is essentially a measure of the degree of
alignment of the charges along the diagonals of the square cell:
\begin{equation}
P=\frac{\rho_1+\rho_3-(\rho_2+\rho_4)}{\rho_1+\rho_2+\rho_3+\rho_4}
\, ,
 \label{1.1}
\end{equation}
where $\rho_i$ is the charge density at site $i$ (see Fig.\
\ref{modelo}a), and the numbering of the sites is in clockwise
order. In previous works, the effect of dissipation on QCA has
been included in the quantum equations of motion in a
phenomenological way, \cite{wang98} through a damping parameter.
This model is shown to give qualitatively correct results only in
the low temperature limit, as it will be evident in our
calculations. Also, the temperature dependence of the cell
polarization has been studied for the case of a cell composed of
two Coulomb-coupled double dots, \cite{mourokh} and the
contribution of relaxation processes to the speed of the cell
response was discussed using stochastic equations for the
population differences and dipole moments of the double dots. The
effects of imperfections have been considered before, with
\cite{zozo99,mrs2000,rojas2002} and without \cite{rojas2000}
dissipation, and found to have strong effects in the dynamical
response, by delaying or slowing down the response of the target
cell, and producing a less than optimal desired polarization.  In
this paper, we include the important effect of the environment
through a microscopic model of electron-phonon interactions.  The
thermal bath is modelled with phonons in equilibrium at a given
temperature and structural asymmetries are introduced via size
variations of the quantum dots.

We should point out that in addition to their relevance as a new
computational paradigm, QCA arrays provide a conceptually simple,
yet intricate and subtle, quantum mechanical system in which to
study fundamental coherence and dissipative processes.  The
parametrical requirements to make practical QCA devices yield a
finite set of two-particle states which can be manipulated to
yield insights into the loss of coherence in these and general
quantum systems.  As such, QCA provide then a model condensed
matter system where dynamics and decoherence can be explored in
detail.  To this effect, we use a quantum open systems approach
\cite{mahler, sunmilburn, mrs2000} to study the dynamical
evolution of the {\it reduced density matrix} $\hat\rho_S(t)$ of
the system, and evaluate the competition between coherent (driver
cell) and incoherent (thermal bath) field effects in the response
of the basic cell. We study the effects of temperature,
electron-phonon coupling intensity, and structural asymmetry on
the behavior of the polarization as a function of time. In
particular, we discuss the asymptotic polarization after switching
events, and once the driver is in a stationary configuration, for
different temperatures and structure parameters. An analysis of
asymptotic fixed points in the resulting dynamics provides
information on the system, as we will explicitly demonstrate and
discuss later on.

A measure of the mixed character of a system described by a
density matrix $\hat\rho$ is provided by the {\it linear entropy},
$s[\hat\rho]=Tr\{\hat\rho-\hat\rho^2\}$. \cite{zureck}  This
quantity, which can be thought of as the lowest order
approximation to the von Neumann entropy, is also called the {\em
idempotency defect} or the {\em purity} of a state.  Notice that
the linear entropy is zero for a pure state, since ${\rm Tr}
\{\hat\rho \} = {\rm Tr} \{ {\hat\rho}^2 \} =1$ in that case.
Non-zero values of $s[\hat\rho]$ provide then a quantitative
measure of non-purity of the state of a system. When monitored in
time, the linear entropy provides a convenient measure of how fast
the loss of quantum purity occurs in a system in contact with a
bath, or the {\em increase} in coherence under appropriate
`pumping' conditions, as we will discuss below. Other uses of
$s[\hat\rho]$ include the study of quantum manifestations of
chaotic behavior in a complex system. \cite{furuya}  We explore
the influence of temperature, electron-phonon coupling intensity
and driver cell switching times on $s[\hat\rho]$, in order to
understand the process of dissipation in the QCA system.

We find in all cases that high dissipation and asymmetry influence
adversely the response of the basic cell, as one would intuitively
anticipate. However, the behavior with temperature is quite
subtle. For example, low temperature results in better switching
and more faithful following of the driver by the target cell.
Higher temperatures eventually produce quenching of the
polarization and a dynamical evolution which basically ignores the
driver.  On the other hand, size defects in the dots result in
lagging of the switching behavior with respect to the driver, and
a faster switching window.  This is produced by the pinning of a
given polarization state by the energetics of the size defect, so
that the driver potential has to apply a strong force before the
target cell can follow.  Similar lag and/or pinning, as well as
sudden switch of the polarization is seen in the case of two
target cells being driven by a single cell.  A given cell in the
middle of a chain would clearly have to deal with the influence of
more than one cell, resulting in a more complex potential
landscape and corresponding dynamical behavior.  We also find for
a single cell that in the case of low temperature or weak
dissipation the cell {\em recovers} dynamical coherence, due to
the `pumping' of the driver, clearly overpowering the thermal
bath. Other regimes will be discussed in detail.

The behavior of the polarization of the basic cell as a function
of time when the polarization of the driver is fixed, as well as
when it changes linearly, has been briefly discussed recently.
\cite{mrs2000} There, we studied the symmetric (no defect) case,
but taking account of dissipation in the system.  In what follows,
we present a full report of the QCA dynamical behavior under
different electron-phonon and temperature conditions, and analyze
the linear entropy of the system during these processes.

The paper is organized as follows: in section \ref{sec:model} we
describe the model for the entire system, composed of the QCA and
the thermal bath. We use an extended Hubbard Hamiltonian to
describe the Coulomb interactions present in the system (target
cell(s) + driver cell). The thermal bath is represented by a  set
of harmonic oscillators, and the electron-phonon interaction
enters as emission and absorption of phonons during electron
tunneling events between quantum dots in the system.  A Markovian
approach to the dynamical evolution of the reduced density matrix
is presented in \ref{sub:mark}, where the transition rates for the
electron-phonon interaction in this system are explicitly
obtained. In section \ref{sec:res} we present and discuss the
behavior of the polarization and linear entropy, as function of
relevant parameters, with special emphasis on effects of
dissipation and asymmetry. Here, we also discuss the dynamics of
multiple cells, as the QCA ultimate devices involve multiple cell
geometries. Finally, we summarize and discuss our results in
section \ref{sec:conclu}.

\section {Model}
\label{sec:model} Consider a problem described by the Hamiltonian
$\hat H = \hat H_S + \hat H_R + \hat V_{SR}$, where the subscripts
$S$ and $R$ stand for system and reservoir, respectively. In our
case, $\hat H_S$ describes the target cell in the presence of a
driver cell which provides external driving fields on the system.
The interactions are introduced via an extended Hubbard
Hamiltonian with both intra- and inter-cell Coulomb repulsion
terms, as well as intra-cell tunneling,
 \begin{eqnarray}
\hat H_S &=&\sum_j \epsilon_j \hat n_j - \sum_{<ij>} (t_{ij}
 \hat c_i^\dagger \hat c_j + c.c.) \nonumber \\
 &+& \sum_{i>j} V_{ij}\hat n_i \hat n_j
 + \sum_{i,j} W_{ij} n_i^d(t) \hat n_{j} \, .
\label{h-eq}
 \end{eqnarray}
 The first three terms describe the {\em target} or active cell
and the last term describes the Coulomb interaction between the
electrons in the target cell and a {\em driver} cell which is
externally controlled.  Here, $\epsilon_j$ is the confinement
energy level in the $j^{th}$ quantum dot (QD) in the target cell.
In this term, possible dot asymmetries or local potential
variations are introduced by changing the energy in a given
$k^{th}$ QD, $ \epsilon_k \rightarrow \epsilon -\delta$, where
$\delta$ is the measure of the imperfection due to different dot
size or a change in the local environment, and $\epsilon$ is the
energy of the other (identical) dots. The amplitude $t_{ij}$ is
the tunneling matrix element between nearest neighbors $<ij>$ on
the same cell, $ \hat c_j^\dagger$ and $\hat c_j$ are creation and
annihilation operators for site $j$, $\hat n_j=\hat c_j^\dagger
\hat c_j$ is the electron number operator at site $j$, $V_{ij}$ is
the Coulomb interaction between sites $i$ and $j$ in the target
cell, and $W_{ij}$ is that between site $i$ in the driver cell
(with charge density $n_i^d(t)$, which in general changes in time)
and site $j$ in the target cell.  A generalization to multiple
cells is straightforwardly given by inter-cell interactions.
\cite{rojas2000}  Following the notation of Lent {\em et al}.,
\cite{lent93} the distance between nearest-neighbor dots (length
of the square side) is $a$, while the separation between the
target and driver cells, located side by side, is $c$ (see inset
Fig.\ 1a).  Notice we consider spinless electrons, although
considering two-particle states with given total spin is also
possible in our formalism.

The Hamiltonian for the reservoir is given by a set of harmonic
oscillators of frequency $\omega_k$, $\hat H_R=\sum_k \hbar
\omega_k \hat b_k^\dagger \hat b_k$, where $\hat b_k^\dagger(\hat
b_k)$ are creation (annihilation) operators of phonons. Finally,
the electron-phonon interaction is given by a general expression,
\begin{equation}
\hat V_{SR} = \sum_k\sum_{<ij>}\alpha_{kij}\hat c_i^\dagger \hat
c_{j} (\hat b_k^\dagger + \hat b_k) \, , \label{eq-int}
\end{equation}
corresponding to emission/absorption of a phonon for electron
tunneling events and with matrix elements given by the
coefficients $\alpha_{kij}$. Notice that the physics of the phonon
interactions in the system is well described by the model chosen
here. Although phonons are in principle present everywhere, and
not only during tunneling events, they only provide an `on-site'
energy renormalization/shift and broadening, which will be ignored
here, as it yields a uniform correction to the entire cell.  In
contrast, the phonon-assisted tunneling events described by
$V_{SR}$ affect significantly the dynamics of the cell, as they
provide an effective coupling mechanism among the different levels
in the spectrum and allow for energy relaxation processes.

We assume, as in [\onlinecite{lent}], that the quantum dots have a
characteristic size $d \simeq 50$ nm defined on a GaAs/AlGaAs
heterostructure, with effective mass $m^*=0.067$. The typical
distance between nearest-neighbor dots is $a \gtrsim 100$~nm, and
the dielectric constant of the medium is $\approx 12$. These
values give an estimate of the Coulomb repulsion between nearest
neighbor dots of $V\approx 1$~meV, which is taken as the unit of
energy in this paper. One should notice that the first excited
single-particle state in each dot lies at quite a high energy,
$\Delta \approx (\hbar^2/2m^*) (2 \pi/d)^2 \approx 9$~meV $\gg V$,
so that one can safely consider a single orbital per site,
although additional single-particle levels in each dot could in
principle be included.  Notice also that the on-site Coulomb
energy is quite high and prevents the double occupation of the
dot, so that our basis set of states ignores that possibility. We
should comment that successful implementations of QCA elements
using {\em metallic} islands are in a multiple-orbital regime,
since in that case $\Delta \ll V$ for typical 1$\mu$m-size
islands.  The condition of overall electrical neutrality of the
target (and driver) cell is included in the calculation by
distributing a uniform positive background charge $+2e$ throughout
the cell ($e/2$ in each dot), which results in multipolar fields
from/at each cell.

Notice that we study the behavior of the target cell interacting
with the driver via Coulomb interaction only (no inter-cell
tunneling allowed). This interaction is described by the last term
in the Hamiltonian $H_S$, Eq.\ (\ref{h-eq}), and assumed to be
time dependent via $n_i^d(t)$. For a fixed driver and no
dissipation, the solution can be found by direct numerical
diagonalization of $H_S$. In that case, the effect of various
parameters on the polarization have been studied to show that the
energy spectrum is strongly influenced by even small defects,
which affect the bistability of the cell. \cite{felipe}  In the
case of a time-dependent driver, we have to take into account that
the driver promotes transitions to intermediate excited states. In
fact, the dynamical competition between the driver (coherent terms
in the equation of motion) and the bath (dissipative terms) causes
a non-trivial time evolution of the system. We have studied
different `driving schemes' but will mostly consider here the case
of a driver with a linear switching of its polarization,
$P_{driver}(t) = 1-2t/ \tau$, for times $ 0 \leq t \leq \tau$,
where $\tau$ is the switching time. This corresponds to the
configuration of charge densities in the driver cell of
$n_1^d(t)=n_3^d(t)=1-t/\tau$ and $n_2^d(t)=n_4^d(t)=t/\tau$. For
the four-dot cell arranged on a square, we take $V_{ij}=V$ for
nearest-neighbor dots, while all others scale with their (inverse)
separation.  We also set all $t_{ij}$ equal to a constant $t$
between nearest neighbors only (not to be confused with time). \\

\subsection{Markovian master equation}
\label{sub:mark}

In the interaction picture ($I$), the evolution of the total
density matrix is given by
 \begin{eqnarray}
\frac{\partial \hat \rho^{(I)}(t)}{ \partial t} &=& -
\frac{i}{\hbar} \left[ \hat V_{SR}^{(I)}(t),\hat \rho^{(I)}(0)
\right] \nonumber \\
&-& \frac{1}{\hbar^2} \int^t_0 dt' \left[ \hat V_{SR}^{(I)}(t),
[\hat V_{SR}^{(I)}(t'),\hat \rho^{(I)}(t')]\right] ,
 \label{eq-dentot}
\end{eqnarray}
where $\hat V_{SR}^{(I)}$ is the cell-bath interaction operator,
and the expansion is valid up to second order in $\hat
V_{SR}^{(I)}$. \cite{mahler}

Our approach uses a closed and time-local equation for the {\it
reduced density matrix} (RDM): ${\hat{\rho}_S}^{(I)}(t)= {\rm
Tr}_R \{\hat{\rho}^{(I)}(t)\} $ where the trace is carried out
over the degrees of freedom of the reservoir, $R$. The following
fundamental assumptions are made in this approach: (1) the system
$S$ and reservoir $R$ are initially uncorrelated, which implies
$\hat\rho(0)= \hat \rho_S(0)\hat\rho_R(0)={\hat\rho }^{(I)}(0)$;
(2) a stationary reservoir at temperature $T$ with equilibrium
density matrix $\hat\rho_R(0) = \exp( -\hat{H}_R/k_B T )/{{\rm Tr}
\{ \exp ( -\hat{H}_R/k_B T )\}}$, exists at all times; and (3) the
correlation time $\tau_c$ in the reservoir is much shorter than
characteristic times for the RDM system to change appreciably,
meaning that ${\hat\rho_S}^{(I)}(t')\approx
{\hat\rho_S}^{(I)}(t)$, if $t'-t \alt \tau_c$, while the state of
the reservoir at time $t'$ has already no correlation with the
state at time $t$ (Markov approximation). Thus one arrives at the
time evolution equation for the RDM elements (details of the
derivation can be found in the literature \cite{mahler, mrs2000}),
which in the Schr\"odinger picture is given by
\begin{equation}
{\dot{\hat\rho}_S(t)}_{ss'}=-i\omega_{ss'}{\hat\rho_S (t)}_{ss'}+
\sum_{mn}\tilde R_{ss'mn} \, {\hat\rho_S (t)}_{mn} \, .
 \label{eq-master}
\end{equation}
 The first term on the right represents reversible dynamics
(``coherent" effects), in terms of the transition frequencies
$\omega_{ss'}=(E_{s}-E_{s'})/\hbar$, while the second term
describes relaxation (irreversible dynamics). In this part, the
{\it relaxation tensor} $\tilde R_{ss'mn}$ can be written
explicitly as
\begin{widetext}
\begin{equation}
\tilde{R}_{ss'mn}=\left \{ \begin{array}{ll}
\delta_{nm}(1-\delta_{ms}) \tilde{W}_{sm}-\delta_{ms}
\delta_{ns}\,
\sum_{k\neq s} \tilde{W}_{ks} , & (s=s')\\
-\gamma_{ss'}\, \delta_{m s} \delta_{n s'}, & (s\neq s')
\end{array}
\right.
\label{eq-Rtilde}
\end{equation}
\end{widetext}
Here, $\tilde{W}_{nm}$ are {\it transition rates} from state
$|m\rangle$ to $|n \rangle$, which in terms of the properties of
bath and target cell, can be  expressed as
 \begin{eqnarray}
 \tilde{W}_{mn} &=&\!\! \frac{2 \pi}{\hbar^2}D^2 |g(\omega_{mn})|^2
 {\mathcal D}(\omega_{mn}) \left\{ |S_{nm}|^2 \bar{n}(\omega_{mn})
 \right.  \nonumber \\
 &+& \left. |S_{mn}|^2 [1\!+\!\bar{n}(\omega_{mn})] \right\}
  \label{eq-trans}
 \end{eqnarray}
where $S_{mk}$ are matrix elements of the electronic part of the
target cell interacting with the phonon reservoir, Eq.\
(\ref{eq-int}); $\bar{n}(\omega)=\langle
\hat{b}^\dag(\omega)\hat{b}(\omega)\rangle =
(e^{\beta\hbar\omega}-1)^{-1} $ is the average phonon number in
the reservoir at temperature $T=1/k_B \beta$.  The first term
($\propto \bar n$) describes the absorption of phonons, while the
second ($\propto \bar n +1$) describes phonon emission, both at a
frequency given by the system's transition energy $\omega _{mn}$.
Notice that the matrix elements $S_{mn}$ embody the appropriate
selection rules for the different states involved in the
transitions. For simplicity, we consider a model where the
amplitude of interaction does not depend on the electronic states
of the target cell, and has the form $\alpha_{kij}= Dg_k$, where
$D$ is the constant prefactor of the electron-phonon interaction
and is determined from the deformation potential model.
\cite{D-note} We should emphasize that this model introduces
dissipation effects only at finite temperature, and cannot account
for elastic decoherence effects. \cite{Caldeira-Leggett} Here,
$g_k=g(\omega_k)\sim \omega_k^{1/2}$ describes the frequency
dependent amplitude of the electron-phonon interaction in the
deformation potential model; $\mathcal D(\omega)\sim \omega^2 $ is
the density of states in a Debye model and $\omega_{mn}$ is the
transition frequency between states $|m \rangle$ and $|n \rangle$.
It is clear that other phonon channels with different dispersion
relations and/or matrix elements exist, but we believe that other
modes would not be as effective in introducing dissipation in this
system (e.g., optical phonons occur at energies too high to couple
effectively), nor qualitatively change the results discussed here.
In Eq.\ (\ref{eq-Rtilde}), $\gamma_{ss'}$ is a so-called
non-adiabatic parameter whose real part gives a contribution to
the time decay of the off-diagonal density matrix elements, and is
then directly responsible for the loss of coherence. These
parameters can be written in terms of the transition rates by
${\rm Re}~\gamma_{ss'} = (\sum_{k\neq s} \tilde{W}_{ks}+%
\sum_{k\neq s'} \tilde{W}_{ks'})/2$.  The imaginary part of
$\gamma _{ss'}$ is an intrinsic relaxation rate for each
transition, which we assume negligible, ${\rm Im}~\gamma_{ss'}=0$,
as it could be incorporated in the transition frequencies,
$\omega_{ss'}$. \cite{mahler}

It is interesting to point out that Eq.\ (\ref{eq-master}),
written for the diagonal elements, takes the form:
\begin{equation}
 {\dot{\hat\rho}_S(t)}_{ss}=\sum_m
 \tilde{W}_{sm} \, {\hat\rho_S(t)}_{mm} - {\hat\rho_S(t)}_{ss}
 \sum_m \tilde{W}_{ms} \, ,
\end{equation}
which has a simple interpretation: the probability of finding the
level $|s \rangle$ occupied at time $t$,
${{{\hat\rho}_S}(t)}_{ss}$, increases due to transitions from all
other levels $|m \rangle$ {\em to} $|s \rangle$ (first term) and
decreases due to transitions {\em from} $|s\rangle$ to all other
levels $|m \rangle$ (second term).

Note also that the condition ${\dot {\hat\rho}_S(t)}_{ss'}=0$ in
Eq.\ (\ref{eq-master}) corresponds to `fixed points' of the
dynamics of the system, as both the terms related to coherence,
$(\hat\rho_S)_{ss'}$ for $s \neq s'$, as well as the diagonal
terms which determine the population in each state, do not change
in time. These fixed point values depend only on the properties of
the transition rates and the energy spectrum of the system (as the
external drive reaches a stationary configuration), and provide a
direct determination of the dissipation times produced by the
thermal bath. It can be seen from Eq.\ (\ref{eq-trans}) that the
transition rates satisfy
\begin{equation}
\frac{\tilde{W}_{nm}}{\tilde{W}_{mn}}=\exp
\left(-\frac{\hbar\omega_{nm}}{k_B T} \right) \, ,
 \label{eq-cociente}
\end{equation}
which guarantees the detailed balance condition. It is important
to emphasize that we have derived an explicit expression for the
transition rates $\tilde{W}_{mn}$, which would behave at low
temperatures, $T \rightarrow 0$, and/or low dissipation (small
$D^2$) as the form used by Zozoulenko {\em et al.}, in a
qualitative way. \cite{zozo99} Notice however that the
normalization condition in [\onlinecite{zozo99}] is an added
requirement in their phenomenological approach, while it is
included naturally in our treatment. We should also mention that
the detailed balance ratio in (\ref{eq-cociente}) will determine
the fixed point values of $\hat\rho_S$ and all the corresponding
physical properties, including the different state populations.
Since these will in general be asymmetric, the polarization value
of the QCA is reduced, and as we will see later, this results also
in a limiting value for the linear entropy. A detailed analysis of
the asymptotic fixed points in terms of the structure parameters
will also shed light into the dynamical evolution of the system,
as we will discuss.

The solution of the master equation (\ref{eq-master})  for the
time evolution of the RDM is carried out numerically using a
Runge-Kutta fourth order algorithm.  We ensure, by a proper choice
of the integration step, that the general normalization property
${\rm Tr} \{{\hat\rho}_S \}=1 $ for the RDM is satisfied at all
times. Notice that an instantaneous basis is incorporated in the
calculation to take into account the evolution of the ground state
induced by the time-dependent driver configuration.  The basis
functions are then defined by the eigenvalue/eigenvector problem
of the coherent Hamiltonian of the target cell, $\hat{H}_S(t)
|m\rangle = E^S_m(t)|m\rangle$, for a given time $t$ while the
driver is in the process of switching, and the instantaneous
energy in the target cell is $E^S_m(t)$.  Notice also that the
characteristic time variation of the driver $\tau$ is long
compared with the target-cell's natural frequencies
$\omega_{ll'}$, so that $\tau \gg \hbar/\omega_{ll'}$.  This is
effectively the requirement for an adiabatic switching of the
driver.  As is clear in Fig.\ \ref{modelo}, the characteristic
frequencies of the system are $\omega_{ll'} \alt 1$~meV, so that
$\hbar/\omega_{ll'} \agt 1$~ps.  This would require $\tau$ to be a
few nanoseconds for the adiabatic switching regime to be valid.

The numerical solution of the RDM is used to calculate the
dynamics of the polarization in the target cell, as the physical
observable of interest here. The dynamical behavior of $P(t)$,
calculated from the charge density in each site as $ \rho_i(t)=
{\rm Tr} \{\hat \rho_S(t) \hat n_i \}$, and of the linear entropy
$s[{\hat\rho}_S]$, are then determined for a given set of
parameters.  In order to get better insights in the problem, we
also carry out a calculation of the fully coherent dynamics,
ignoring the phonon thermal bath (setting $D=0$ in Eq.\
\ref{eq-trans}). In the coherent limit, the relevant parameters
for the dynamical evolution are those defining the structure,
including the tunneling amplitude $t$, the separation of target
and driver cell $c$, the switching time of the linear driver
$\tau$, and the size of the dot asymmetry $\delta$, if any. For
the dissipative problem, on the other hand, it is clear that in
addition to the structure features, the important parameters are
the temperature of the thermal bath $T$, and the amplitude of the
electron-phonon interaction $D$. Of course, the phonon density of
states, as well as the interaction details, will also play a role
in the evolution of the system. Our model parameters provide an
adequate and realistic description of these systems in
semiconducting QCA.  \cite{D-note}

\section{Results and Discussion}
\label{sec:res}

Figure \ref{modelo}a shows the level spectrum in the target cell
for a fixed driver polarization, $P_{driver}=1$, as shown in the
inset. We see that the four-fold symmetry of the structure and
consequent degeneracy is broken by the driver, even for $t=0$.
\cite{felipe} Moreover, increasing the intra-cell tunneling $t$
produces additional level splittings, as one would expect.  We
should point out that the polarization of the ground state follows
that of the driver, $P_{cell}^{grnd} \simeq 1$, but increasing
hopping $t$ reduces that value somewhat.  The first excited state
has $P_{cell}^{exc} \simeq -1$, and is separated from the ground
state by a gap of $\simeq 0.2$ in the range shown (our energy
units are $V \simeq 1$~meV). The four higher states have vanishing
polarization, as the charges are located on neighboring sites
along the sides of the square. The basic operation of the QCA
requires adiabatic evolution from a polarized ground state to the
other with opposite polarization. Transitions to higher excited
states allow for non-adiabatic terms which yield to depolarization
of the target cell.  Figure \ref{modelo}b shows the influence of a
``defect" in dot size. Here, the dot at site 1 has an energy lower
than the others by an amount $\delta$ (corresponding to a larger
dot). Increasing this asymmetry produces a strong shift of the
spectrum, and perhaps most important for QCA operation, a change
of the polarization of the first excited state.  This is apparent
in the anticrossing seen in \ref{modelo}b for $\delta \simeq 0.1$,
where an opposite polarization state takes over as the first
excited state. \cite{felipe} The defect then will produce a
pinning of the polarization when operating the QCA if the $\delta$
is small, and even prevent the switching of the target cell
altogether if $\delta$ is large.

\begin{figure}
\includegraphics[width=3.6in]{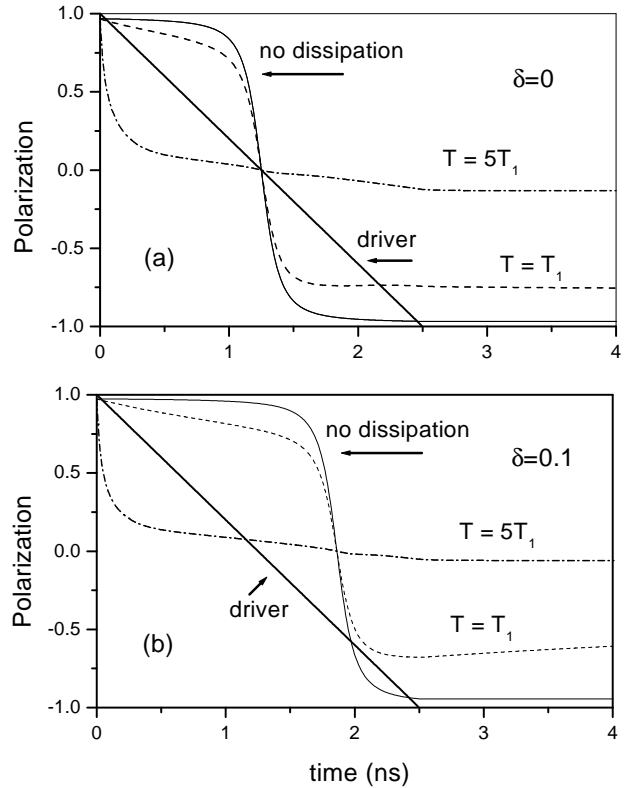}
\caption{Time evolution of target cell polarization as driver cell
switches linearly from $P_{driver}=+1$ to $-1$ in a time $\tau =
2.5$~ns, for different temperatures. (a) Symmetric cell with four
identical dots ($\delta=0$). (b) Asymmetric cell with one lower
energy dot, $\delta =0.1$, at site 1. In all plots, $t=0.03$ and
$D^2=0.05$.} \label{Psym}
\end{figure}

When the polarization of the driver cell changes linearly with
time, the polarization of the target cell evolves as well, as
presented in Fig.\ \ref{Psym}a for the non-dissipative case, and
for different temperatures.  One can see that the polarization of
the target cell follows the driver very well, and switches
smoothly over a short time window ($\approx \tau/5$) centered
about the time when $P_{driver} = 0$.  This general behavior is
not affected for different values of the switching time $\tau$ we
tested (over a wide range within the adiabatic regime; results not
shown). \cite{initial} We see that the response of the target cell
deteriorates for increasing temperature.  At low temperature,
$T=T_1 \simeq 1$~K, the target cell has a slightly decreased
polarization after the driver is fully switched, and the switching
time of the cell is not greatly affected.  However, for $T=5T_1$,
the target cell polarization is quenched after 0.5~ns and ignores
even a fully polarized driver. [Notice that the natural unit of
temperature in this problem is the difference in energy between
the first and second excited states, when the driver is fixed at
$P_{driver}=1$ (Fig.\ \ref{modelo}b), as this transition has one
of the smallest natural frequencies $\omega_{ll'}$, and it
produces depolarization of the cell, as discussed above.  In
particular, we take as characteristic its value for $t=0.03$ in
the symmetric case, $T_1=0.0862 \approx 1$~K. Temperatures are
compared to this natural scale in the problem.] The deterioration
of the response of the cell with temperature is consistent with
Eq.\ (\ref{eq-cociente}) where the different rates at low
temperature would favor transitions to the ground state, while as
temperature increases, transitions up and down become equally
probable.  This results in more symmetric level populations, more
equal electron probabilities in all dots, and consequently the
depolarization of the target cell.  It is somewhat surprising that
even this low temperature would yield such strong quenching of the
QCA polarizability (a value which of course depends also on how
strong $D$ is in Eq.\ \ref{eq-trans}). Figure \ref{Psym}b shows an
example of the effects of asymmetry (quantum dot at site 1 has
lower energy than the rest), for the same parameter values and
temperatures as in Fig.\ \ref{Psym}a. We observe that at low
temperatures, the response of the target cell lags behind the
driver, although the switching time is not affected substantially.
However, the target cell reaches asymptotic values of polarization
which are smaller (in absolute value) than in the ideal symmetric
case. In this case too, changing $\tau$ over a wide range does not
affect this lagging behavior. It is evident that for $T=5T_1$, the
target cell is almost completely depolarized, despite the ``help"
of the larger dot which pins the polarization of the ground and
excited states to $\simeq +1$ (and induces the lag seen at low
temperatures).

\begin{figure}
\includegraphics[width=3.6in]{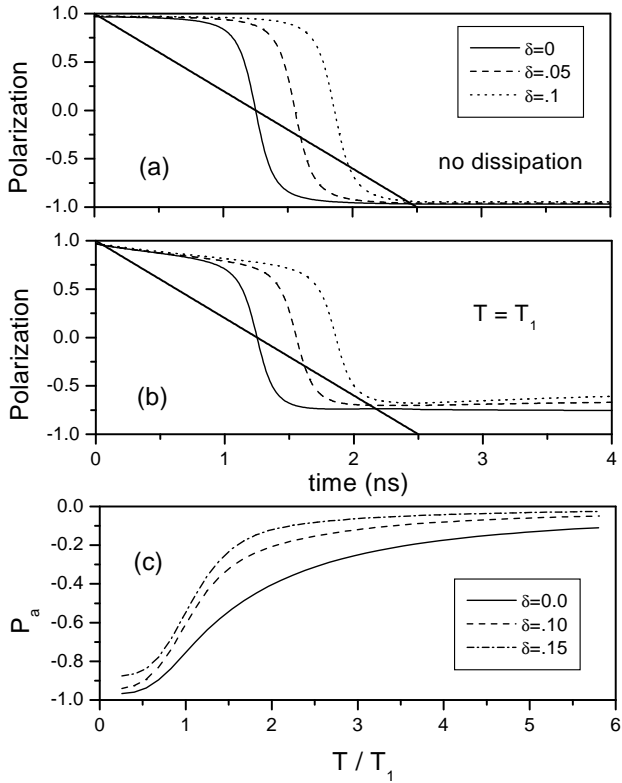}
\caption{Time evolution of target cell polarization with a linear
switch drive, $\tau=2.5$~ns, for different asymmetry $\delta$. (a)
No dissipation ($D=0$); (b) at temperature $T=T_1$. Notice target
cell lagging the switch of the driver increases with $\delta$. (c)
Long-time asymptotic polarization, $P_{a}$ vs.\ temperature $T$
for different $\delta$ values. In all panels, $t=0.03$ and
$D^2=0.05$.} \label{Pasym}
\end{figure}

Figure \ref{Pasym} shows more of the effects of asymmetry and
temperature on the response of the target cell.  Figure
\ref{Pasym}a shows the results for no dissipation, and panel (b)
those for $T=T_1$. From these figures, we see again that the
important effect of asymmetry is to produce a lag or delay in the
response of the target cell as asymmetry increases. On the other
hand, the effect of temperature is to decrease the asymptotic
polarization in absolute value, associated with the delocalization
of electrons within the cell.  Notice also in Fig.\ \ref{Pasym}b
that for finite $T$, the polarization of the target cell continues
evolving in time (deteriorating), even after the driver is fully
polarized. In all cases, the polarization reaches asymptotic
values $P_a$ (long after the driver has stopped changing) which
are a manifestation of `fixed points' of the system dynamics for
the quantum dot populations.  These long-time values of the
polarization depend on temperature (and indirectly on structure
parameters and energy levels) as shown in Fig.\ \ref{Pasym}c, for
different asymmetries. The nearly full depolarization of the
target cell at high temperature is evident and anticipated.  What
is somewhat counterintuitive is the fact that the depolarization
increases with asymmetry, as the pinning that yields the lag
assists now in the depolarization by favoring the population of
the states with the larger dot.

\begin{figure}
\includegraphics[width=3.9in]{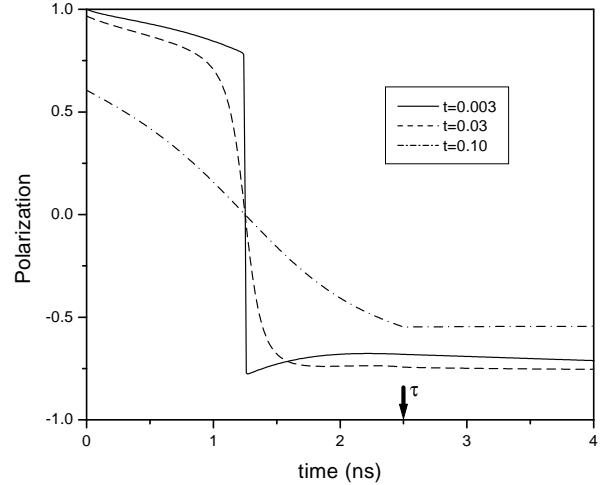}
\caption{Time evolution of polarization in a linear switch driver,
$\tau=2.5$~ns (see arrow).  Increasing tunneling between dots
first improves ($t=0.03$) and then deteriorates ($t=0.1$) the
final polarization. $D^2=0.05$, $T=T_1$, and $\delta=0$ in all
plots. } \label{tunel}
\end{figure}

\begin{figure}
\includegraphics[width=3.6in]{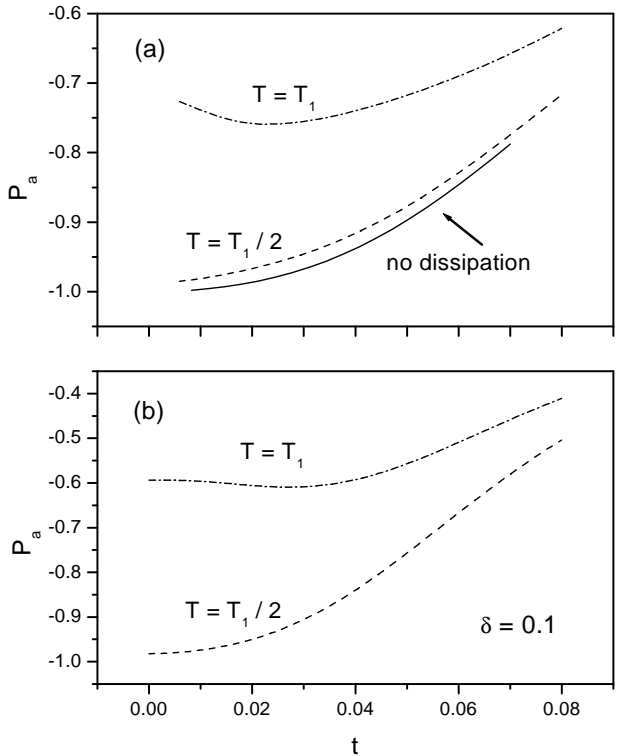}
\caption{Long-time asymptotic polarization $P_a$ vs.\ hopping $t$
at different temperatures, as in Fig.\ \ref{tunel}.  The
non-monotonic behavior is seen only for $T \approx T_1$. (a)
Symmetric cell ($\delta=0$); (b) asymmetric case, $\delta=0.1$. In
both panels, $D^2=0.05$.} \label{Pa}
\end{figure}

Figure \ref{tunel} shows the effects of tunneling coupling
intensity $t$ on the time evolution of the polarization for a
symmetric cell system ($\delta=0$).  We see that increasing
hopping between dots first introduces a smoother transition
following the linear driver ($\tau =2.5$~ns) and a better
polarization.  Increasing $t$ however, delocalizes the charges in
the quantum dots and reduces the polarization of the target cell,
as one can see from the long time limits in the figure.  In order
to look more closely at the non-monotonic behavior of $P_a$, we
analyze this quantity as a function of the tunneling coupling $t$
in Fig.\ \ref{Pa}a. We find that, for $T=T_1$, $P_a$ attains a
minimum value at $t \approx 0.02$. From this figure we can see
that the drop in polarization at small values of $t$ is due to
dissipation processes: We expect that at low $t$ values the
electrons would tend to be well localized at the corners of the
cell (ground state), giving $P=\pm 1$, depending on the driver.
This is indeed what is observed at low temperatures. However, as
temperatures increases, depolarization transitions from the ground
state to excited states become more likely.  As $t$ increases
substantially, the polarization decreases (in absolute value) due
to the delocalizing effects of both tunneling and dissipation, as
expected.  However, for $t \alt 0.02$, the possibility of
transitions which in fact populate the ground state increases, and
therefore improve slightly the asymptotic polarization of the
system. Qualitatively similar results are obtained in the
asymmetric case (Fig.\ \ref{Pa}b), although with lower absolute
values of the asymptotic polarization in this case, as discussed
above (Fig.\ \ref{Psym} and \ref{Pasym}), and a shallower
oscillation.

\begin{figure}
\includegraphics[width=3.6in]{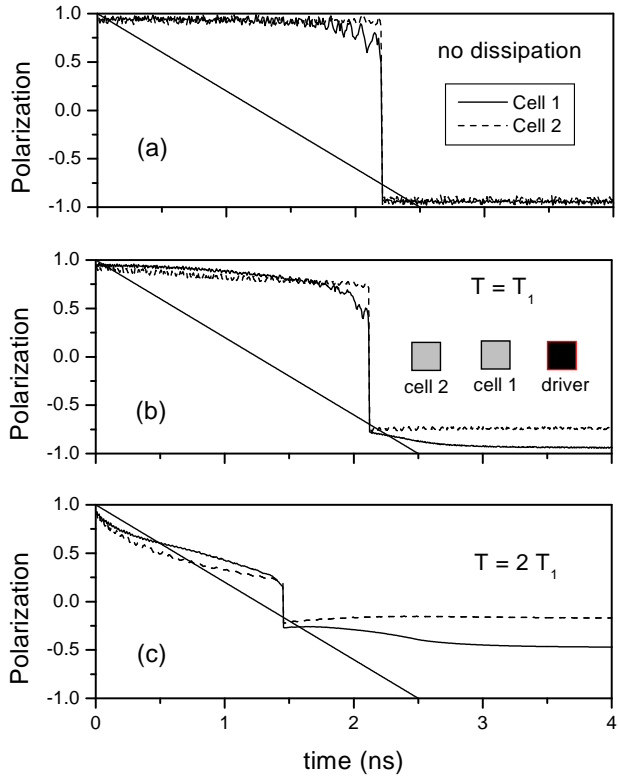}
\caption{Polarization of two active cells and linear driver, as
shown in inset in panel (b). (a) Coherent dynamics; (b) at
$T=T_1$; (c) at $T=2T_1$.  Notice cells lag the driver, except at
higher temperatures. More distant cell (2) switches more suddenly
than cell 1, and has smaller $P_a$.  In all plots, $t=0.03$ and
$D^2=0.05$.}
 \label{2celdas}
\end{figure}

We have also studied the dynamics of two target cells and driver
in line (Fig.\ \ref{2celdas}).  There are a number of interesting
features in this case.  We notice first that the switching for
both target cells lags behind the driver considerably, as the
cells switch only after $P_{driver} \agt 75\%$.  Moreover, the
switch occurs over a much shorter time scale than for the single
target cell ($\simeq \tau /25$).  This change is understandable if
one thinks that the target cell closest to the driver (cell 1) is
in essence between two cells with different polarization and has
its wrong polarization somewhat pinned (this is behavior also seen
in other work \cite{zozo99}).  Increasing temperature to $T=T_1$
makes the switching a bit smoother although still with a sudden
change, and with not as good final polarization.  Notice in
particular that the asymptotic polarization of target cell 2 is
poorer than that of cell 1.  As the temperature increases to
$T=2T_1$, the target cells barely follow the drive, and the
polarization of cell 2 is basically zero at long times.  It is
clear that control of the tunneling process as the ``signal" is
transmitted down the line would be essential in proper operation
of QCA arrays.  Such ``clocked switching" has been discussed by
Lent and coworkers as a solution to steer cells into better
compliance. \cite{lent-tougaw97} Coherent simulations of such
steering,\cite{rojas2000} as well as experiments in metallic QCA
\cite{orlov2000} verify that this is a good operating scheme.

\begin{figure}
\includegraphics[width=3.6in]{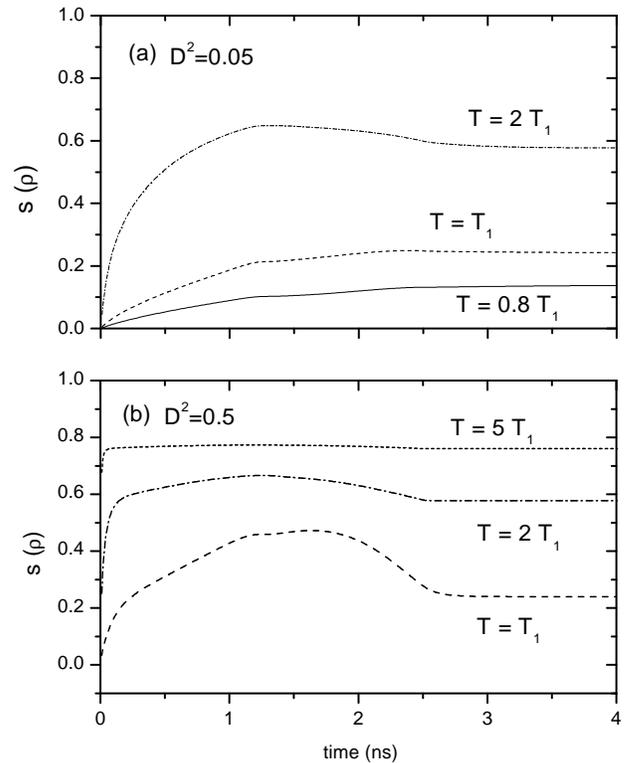}
\caption{Time evolution of linear entropy or purity of the system
in a linear switch, $\tau=2.5$~ns at different temperatures.  (a)
Weak coupling, $D^2=0.05$, $t=0.03$; (b) strong coupling,
$D^2=0.5$, $t=0.03$.  Notice entropy {\em drop} after $\tau/2$ at
intermediate dissipation.} \label{linearentropy}
\end{figure}

Knowing the time evolution of the density matrix we can also
calculate the linear entropy of the QCA system $s[{\hat\rho}_S]$,
to monitor the degree of non-purity introduced during the
switching process by the thermal bath. In Fig.\
\ref{linearentropy} we show the results for different
temperatures, for the case of weak and strong electron-phonon
interactions, as measured by the value of $D^2$ in Eq.\
(\ref{eq-trans}). As in previous cases, the polarization of the
driver cell is switched linearly from $+1$ to $-1$ in a time
$\tau=2.5$~ns. In Fig.\ \ref{linearentropy}a, the case of `weak'
electron-phonon coupling, $D^2=0.05$, we see that at low
temperatures $s[{\hat\rho}_S]$ increases monotonically until it
reaches a final stable value.  It is interesting to see that as
$P_{driver}=0$ or $P_{driver}=-1$ (at 1.25 and 2.5~ns,
respectively), one can see kinks in the entropy curve, indicating
the switch off of driving forces.  The long time asymptotic value
increases (coherence decreases) with increasing temperature, as
one would expect. However, at still higher temperatures and/or for
strong electron-phonon coupling (Fig.\ \ref{linearentropy}b),
$s[{\hat\rho}_S]$ first increases and then decreases. In these
cases, we see a maximum in the linear entropy curve at $\approx
\tau/2$, when the polarization of the driver cell goes through
zero.  Notice in fact that whenever $P_{driver} \simeq 0$, the
level spectrum in the target cell shows a great deal of
degeneracies.  As the driver turns negative (for time after
$\tau/2$) the splittings are reinstated and the effect of the
driver seems to be one of a coherent `pump' which produces a drop
in the purity, even as dissipative transitions are present. In
other words, the thermal dissipation seems to decrease with
respect to the coherent transitions induced by the driver,
producing a drop in the linear entropy of the system. After the
driver is stationary, after time $\tau$, the thermal transitions
quickly produce the steady state in the density matrix (or level
populations) given by the detailed balance equation
(\ref{eq-trans}). For stronger phonon coupling, $D^2=0.5$, and
high temperatures (Fig.\ \ref{linearentropy}b), the purity of the
system deteriorates further and the driver is not able to `push'
the system back into a more coherent state.  Notice that the state
of `maximal delocalization' would be achieved as $\hat\rho_{Sii}
\simeq 1/n$, where $n$ is the size of the basis, so that
Tr$\{\hat\rho^2 \} \simeq 1/n$.  In that case, one would obtain $s
\simeq 1 - 1/n$, giving 5/6 in our case.  This value is still
higher than the asymptotic value of $s$ for the highest
temperatures in Fig.\ \ref{linearentropy}b, so that even then the
system is not `fully' decoherent.

\begin{figure}
\includegraphics[width=3.6in]{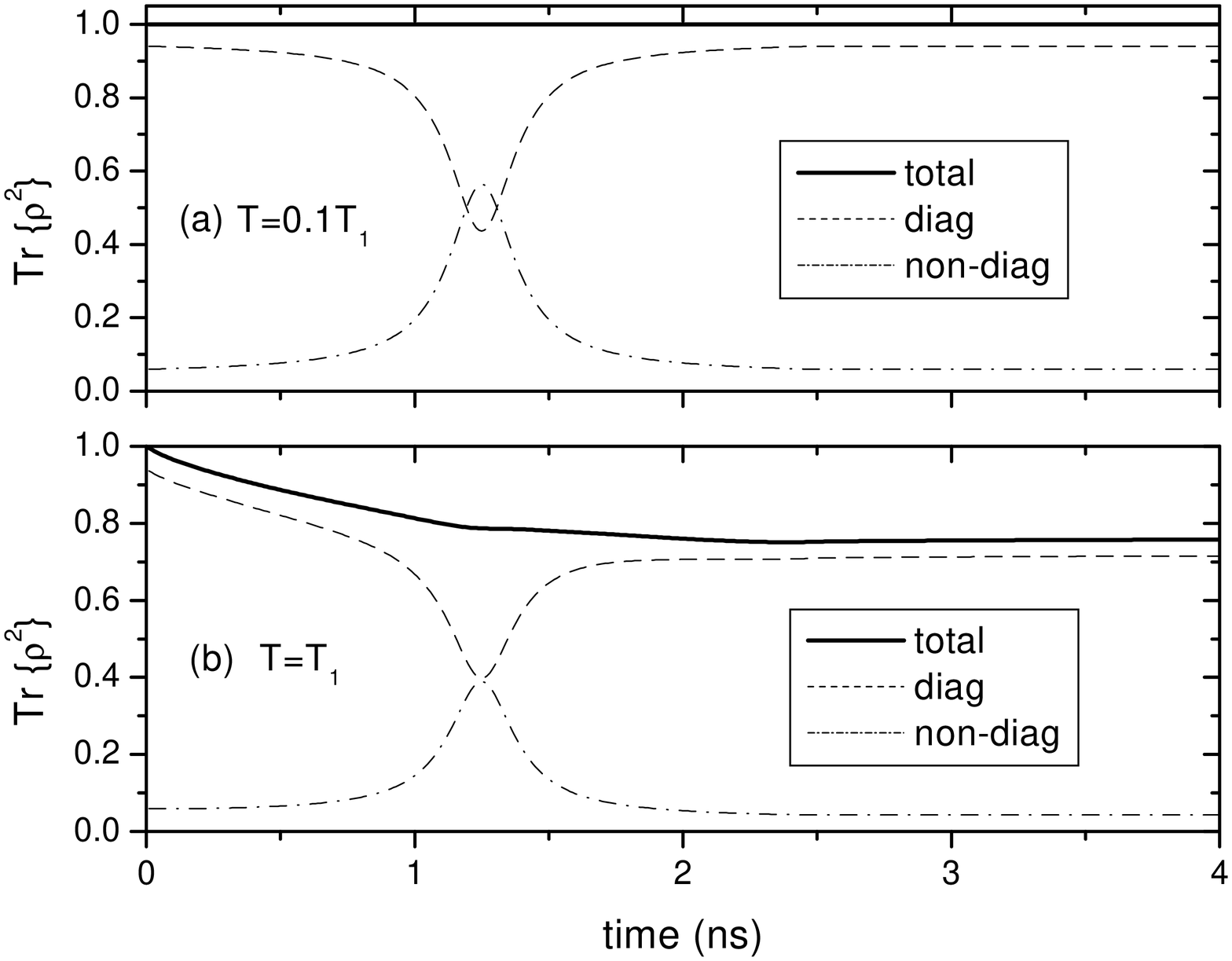}
\caption{Trace of $\hat\rho^2$ vs.\ time during a linear switch of
the driver cell, $\tau=2.5$~ns, and different temperatures.  (a)
$T=T_1/10$, (b) $T=T_1$.  Both diagonal and non-diagonal sums are
shown, as well as the total trace, as in Eq.\ \ref{diag-non-diag}.
In both panels, $t=0.03$ and $D^2=0.05$. }
 \label{traza}
\end{figure}

Let us explore this competition between dissipation and coherent
evolution in more detail.  Since ${\rm Tr} \{\hat\rho_S\}=1$ is a
general property of the density matrix in any representation, one
can instead investigate the behavior of ${\rm Tr} \{\hat\rho_S^2\}
= 1- s[\hat\rho_S]$ directly as a function of the relevant
parameters. \cite{Blum} Notice in fact that this trace has two
contributions
 \begin{eqnarray}
 {\rm Tr} \hat\rho_S ^2 &=& \sum_i (\hat\rho_S^2)_{ii} \nonumber \\
 &=& \sum_i ({\hat\rho_{Sii}})^2 + \sum_{i\neq j}
 |{\hat\rho_{Sij}}|^2
 \, ,
 \label{diag-non-diag}
 \end{eqnarray}
namely, the diagonal and non-diagonal elements of the density
matrix.  Comparing the two terms as a function of time, for
example, would give one further insights of the overall complex
processes, since the behavior of the non-diagonal elements gives
information on the evolution of coherence in the system. Figure
\ref{traza} shows the trace as function of time for two values of
the temperature, explicitly showing the contributions from the
diagonal and non-diagonal elements of $\hat\rho_S$. We see in
Fig.\ \ref{traza}a an increase in the contribution from the
non-diagonal (``coherent") elements, with maximum at $\tau/2$,
exactly compensated by a decrease in the contribution from the
diagonal elements, as the driver and the bath are promoting
transitions.  Notice also that the values of the diagonal
components before and after $\tau/2$ are approximately the same
(with a similar behavior for the non-diagonal terms).  Thus for
low temperatures $T=T_1/10$ (panel a), the sum remains
approximately constant, equal to $1$, and the system is close to
the pure limit with $s\approx 0$. However, for $T=T_1$, we see
that this sum decreases with time, due essentially to the
corresponding decrease in the diagonal elements.  Notice that as
the driver switches and `pumps' the target cell system, the
off-diagonal coherent terms contribute less to the sum as the
temperature increases (its peak value is smaller in (b) than in
(a); so that the system is less coherent indeed), and their
contribution for longer times is lower than at the start of the
switch (time zero). This gradual decay in the coherent
contribution shows even more substantially (and less intuitively)
in the diagonal terms, which one would expect to be less affected
perhaps.  Since Tr$\{\hat\rho \}=1$, the drop in the diagonal
terms of Tr$\{\hat\rho^2\}$ means that the population of each
level (${\hat\rho_{Sii}}$) has been re-distributed and made
somewhat more equal (although still much higher than the `maximal
delocalization' state discussed above with Tr$\{\hat\rho^2 \}
\simeq 1/6$).

\begin{figure}
\includegraphics[width=3.6in]{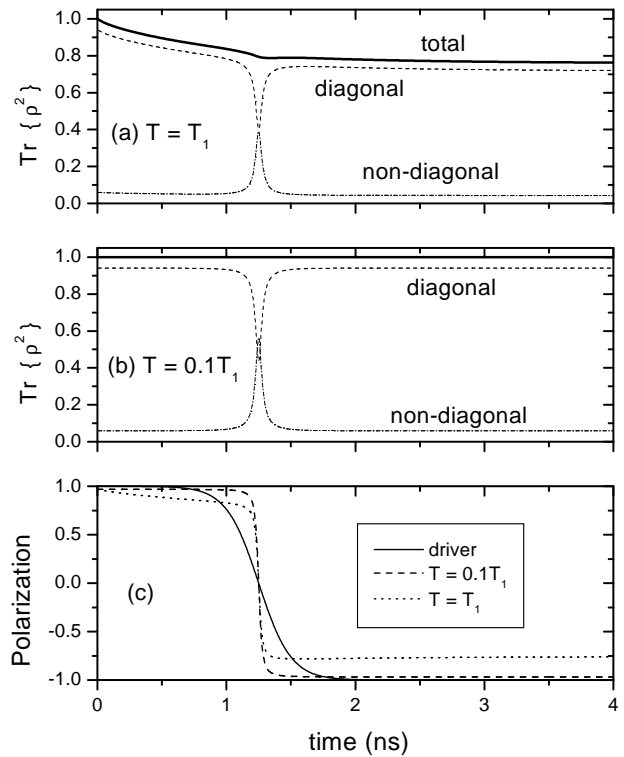}
\caption{Time evolution for a smoother driver switch.  (a) Trace
of $\hat\rho^2_S$ at intermediate temperature, $T=T_1$, showing
diagonal and non-diagonal contributions. (b) Trace contributions
for $T=T_1/10$. (c) Driver (solid curve) and target cell
polarizations for different temperatures. In all panels, $t=0.03$
and $D^2=0.05$.} \label{otrasformas}
\end{figure}

It is also interesting to investigate the influence of switching
time and of the function controlling the switching.  Figure
\ref{otrasformas} shows results for ${\rm Tr} \{\hat\rho_S^2\}$
for a smoother driver transition.  In this case, the driver has no
rapid off/onsets and the evolution of $P_{driver}$ is more
adiabatic near the ends of the sweep than before (see panel c).
Panel (b) shows the trace at low temperatures, which shows a
similar rise (drop) in non-diagonal (diagonal) density matrix
elements as in Fig.\ \ref{traza}a.  However, as the time during
which $P_{driver} \simeq 0$ is shorter now (effectively larger
slope in panel c), the window of variation is also reduced. Higher
temperature in panel \ref{otrasformas}a shows also a narrower time
window variation, but an identical amplitude as in Fig.\
\ref{traza}b.  Other runs with different driver switching slopes
follow this behavior too: narrower windows for larger slopes near
$P_{driver} \simeq 0$ (not shown).  This would suggest that the
amplitude variation of the diagonal/off-diagonal components is
only a function of temperature, while the time window over which
they vary is a function of how long the driver is close to zero
polarization.

\section{Conclusions}
\label{sec:conclu}

We have studied the behavior of the polarization of a square array
of quantum dots with two mobile electrons.  The quantum dynamical
evolution in response to a driver cell whose polarization is
changing with time in a prescribed manner, has been shown to
depend strongly on switching or driver characteristics, symmetry
of the cell and on temperature. Our results show that the
asymptotic values of the polarization in a stationary driver are
fixed points of the system that change with temperature, tunneling
probabilities and imperfections, but not on driving schemes. The
optimal response of the QCA arrays is affected adversely by
temperature and imperfections, with a strong tendency toward
depolarization as temperature or asymmetry increase.  Our
calculations indicate that good control of the QCA is only
achieved for temperatures of at most a few kelvins.  From our two
target cell results, it is likely that ``clocked switching" would
be required for faithful operation of a chain, so that the
polarization information is not degraded down the line. We have
also analyzed dissipation effects on the response of the basic
cell through calculation of the linear entropy or purity of the
system. We find that the study of this quantity is a good tool in
furthering the understanding of a quantum system as relatively
simple, but still quite subtle, as the QCA architecture.  Apart
from providing insights as a function of time and driver
characteristics, the entropy also shows clearly that at high
temperatures, the basic cell is no longer in the required/desired
final state but in one where the polarization tends to zero and
maximizes the entropy.

\acknowledgements

This work was supported in part by CONACYT Project 27702--E.\@ SEU
is supported in part by US DOE grant no.\ DE--FG02--91ER45334.  We
acknowledge support from the Condensed Matter and Surface Sciences
Program at Ohio University.


\begin{thebibliography}{99}
\bibitem{D-note} The value of $D$ in Eq.\ (\ref{eq-trans})
depends on the deformation potential of the semiconductor, as we
consider coupling to acoustic phonons. Typical values of 1 to 3~eV
yield a value of $D^2 \approx 7$ to 20 in our units. As the
potential confinement of the electrons in the dot yields poorer
coupling to acoustic phonons in the solid with characteristic long
wavelengths, the effective value of $D$ is likely much smaller. We
anticipate that the value of $D$ is smaller by a factor of
$\approx 5$, and then use a value of $D^2=0.5$ as our `strong
coupling' regime.
\bibitem{initial} In all the time evolution plots shown, the initial
condition of the target cell is chosen as the ground state of the
system at the given hopping $t$, for $P_{driver}=+1$.  This yields
a smooth non-oscillatory oscillations at the start of the driver
switchings.  Other initial configurations with polarization
similar to the ground state yield time evolutions with similar
characteristics in $P$ and in linear entropy.
\end{thebibliography}
\end{document}